\documentclass[final,1p,times]{elsarticle}

\usepackage{graphics,amssymb,amsfonts, multirow,booktabs,amsmath,hyperref,epsfig,color,eurosym, amstext,color,dcolumn,adjustbox,lscape, graphicx, caption, appendix}
\usepackage{endnotes}
\usepackage{graphicx}
\usepackage{makecell}
\usepackage{caption}
\usepackage{setspace}
\usepackage[utf8]{inputenc}
\usepackage[singlelinecheck=false]{caption}
\biboptions{authoryear}
\graphicspath{{figure/}}
\journal{arXiv}
\DeclareCaptionFont{tenpt}{\fontsize{10pt}{11pt}\selectfont #1}
\begin{document}
\begin{frontmatter}
\title{Close to Home: Analyzing Urban Consumer Behavior and Consumption Space in Seoul}

\author[a]{Hyoji Choi}
\author[b,c]{Frank Neffke}
\author[a,d,e]{Donghyeon Yu }
\author[a,e,f]{Bogang Jun \corref{cor1}}

\cortext[cor1]{\emph{Corresponding}: bogang.jun@inha.ac.kr}

\address[a]{Reaserch Center for Small Business Ecosystem, Inha University, Incheon, South Korea}
\address[b]{Complexity Science Hub Vienna, Vienna, Austria}
\address[c]{Center for International Development, Harvard University, Cambridge MA, USA}
\address[d]{Department of Statistics, Inha University, Incheon, South Korea}
\address[e]{Department of Data Science, Inha University, Incheon, South Korea}
\address[f]{Department of Economics, Inha University, Incheon, South Korea}

\begin{abstract} 

This study explores how the relatedness density of amenities influences consumer buying patterns, focusing on multi-purpose shopping preferences. Using Seoul's credit card data from 2018 to 2023, we find a clear preference for shopping at amenities close to consumers' residences, particularly for trips within a 2 km radius, where relatedness density significantly influences purchasing decisions. The COVID-19 pandemic initially reduced this effect at shorter distances but rebounded in 2023,  suggesting a resilient return to pre-pandemic patterns, which vary over regions. Our findings highlight the resilience of local shopping preferences despite economic disruptions, underscoring the importance of amenity-relatedness in urban consumer behavior.

\end{abstract}
\end{frontmatter}

\section{Introduction}

Recently, economic resilience has been widely examined in various research streams to understand the dynamism of regional economies~\citep{Martin2015notion, Hassink2010resilience, hu2017exploring}, especially in times of crises ~\citep{tooze2021shutdown, Tooze2022Polycrisis}. It is frequently employed to examine the systemic response following an external shock and refers to the idea that systems, like sectors or regional economies, have the capabilities to recover from shocks or can enhance their capacities to manage potential risks ahead~\citep{Martin2015notion, martin2020regional,fromhold2015sectoral,sutton2022regional}. Resilience, in this context, is not just a measure of recovery to a previous state, but an indication of an economy's ability to adapt, transform, and ultimately, prosper post-shock ~\citep{Martin2015notion, martin2020regional}. 

In the realm of economic resilience, the COVID-19 pandemic has cast a new spotlight on consumer behavior patterns. This is because the restricted mobility during the COVID pandemic led to widespread changes across the entire consumption sector. For example, the increase in online shopping has led to changes in consumer shopping trip patterns, and the increased investment in health and wellness suggests the possibility of structural changes in consumption. In line with this, the concept of resilience has evolved to underscore the adaptability and robustness of human agency in the face of crises~\citep{bristow2014regional, kurikka2021resilience}. 

Yet, despite the critical role of human behavior in the adaptive processes that define economic resilience ~\citep{bristow2014regional, kurikka2021resilience}, empirical research on how individual and collective behaviors change in response to such unprecedented disruptions remains sparse. The importance of understanding how consumer behavior adapts during crises is not only academic but also practical, offering crucial insights for policymakers. While extensive discussions on regional economic resilience have been prevalent, the behavioral dimension of this resilience - specifically, the actions and reactions of consumers - has not been sufficiently explored. In this context, this study aims to contribute to the burgeoning discourse on economic resilience by examining the nature of the behavioral response in Seoul to the COVID-19 pandemic, utilizing the economic complexity approach, particularly the concept of relatedness. 

The principle of relatedness, as introduced and developed in scholarly works by \cite{hidalgo2018principle} and further explored by \cite{Balland2019smart}, \cite{boschma2023evolutionary}, and \cite{hidalgo2021economic}, has emerged as a pivotal concept in understanding the dynamics of regional economic development. At its core, the principle posits that the growth potential of an activity within a location can be significantly influenced by the presence of related activities. This notion is not just of academic interest but has profound implications for economic policy-making, serving as a foundational element for place-based development strategies.

Research in this area has typically been applied and studied in the context of the production of products and knowledge at the levels of technology, industry, and national units. However, recent studies, such as those by \cite{hidalgo2020amenity} and \cite{jun2022economic}, have explored the co-location patterns of amenities across U.S. cities and Seoul, respectively, largely based on the location data of stores. In this context, we suggest that there has been a notable gap in research regarding the structural characteristics of consumption behavioral patterns, particularly regarding actual consumption data. Therefore, this study aims to bridge this gap by using a rich dataset provided by BC Card and Korean Small Enterprise and Market Service data to uncover the consumption structural patterns of representative consumer groups through the lens of relatedness, focusing on how the pandemic has influenced these patterns in terms of economic resilience. 

In addition, our methodology involves identifying amenity clusters in Seoul and constructing a `consumption space' that reflects multi-purpose shopping behaviors. This approach allows us to examine the impact of relatedness density—the measure calculated by the proximity of different types of amenities—and how it explains consumer decisions to engage in multi-purpose shopping trips. Through regression analyses and the plotting of effects, we uncover significant findings that demonstrate a pronounced effect of relatedness density on consumer purchases at shorter distances. This indicates a strong preference for multi-purpose shopping trips within close proximity to consumers' residences, a trend that diminishes with increasing distance from amenities. However, intriguingly, the recovery period in 2023 exhibited varying patterns across different shopping distances. This suggests that the impact of relatedness density on sales count, particularly for shopping close to residence, seems to be showing signs of resurgence. 

In summary, our contributions to the literature on economic geography and economic complexity are as follows. Firstly, by integrating the concept of relatedness into the analysis of consumer behavior during the pandemic, this study not only enhances our understanding of economic adaptability and transformation but also offers insights that could inform more resilient economic policy frameworks in the face of future shocks. In addition, given the challenges in accessing detailed consumption data, while much research relies on surveys of self-reported consumer behaviors \citep{chenarides2021food}, we utilize card transaction data that directly reflect consumption activities. Secondly, by calculating relatedness for fine-grained consumption activities, we extend the literature on the Economic Complexity Approach and open a new window in the discussion of relatedness within the consumption sector. 

The rest of this paper is organized as follows: Section 2 provides a literature review on the concept of resilience for consumer behavior and relatedness, Section 3 summarises the data and methods, and Section 4 presents the three main results of this work. Finally, Section 5 concludes and discusses the ideas examined in this paper.

\section{Literature Review}
\subsection{consumer behavior and economic resilience}
Resilience has emerged as a pivotal concept in regional studies, essential for understanding how different local and regional economies respond to economic challenges~\citep{pendall2010resilience, christopherson2010regional, hill2012economic, martin2020regional}. It serves as a lens to explore why under the same economic shock, some regions rejuvenate while others stagnate or decline, addressing a critical question in economic geography regarding the determinants of regional capacity for self-renewal \citep{Hassink2010resilience}.


Academic discussions on regional economic resilience divide into two primary interpretations~\citep{bristow2014regional, martin2007complexity}. The first, from an engineering standpoint, focuses on a system's ability to resist disturbances and quickly return to its original state post-shock, associating resilience with rapid recovery to pre-disturbance conditions~\citep{hill2012economic}. The second, based on complex adaptive systems theory, emphasizes the complex, nonlinear dynamics that enable systems to adapt and reconfigure internally in response to shocks~\citep{Martin2012regional}. This adaptive approach suggests resilience involves not just recovery, but also resistance, restructuring, and growth renewal, offering a deeper understanding of regional economic evolution. The transformative aspect of resilience, advocating for ``bouncing forward'' rather than merely bouncing back, has gained attention for its potential to facilitate progress and enhance future crisis resistance~\citep{giovannini2020time, fromhold2015sectoral, Martin2015notion}. This view proposes analyzing resilience through four stages: risk/vulnerability pre-shock, resistance during the shock, followed by reorientation and recovery, thereby broadening the scope of resilience beyond immediate recovery to include adaptive change and growth post-crisis~\citep{martin2020regional}.

In this context, research into the behavior of transformative agencies is crucial for understanding the mechanisms underlying socio-technical transformations or reorientations following crises. ~\cite{bristow2014regional} highlighted the vital role that human agency plays in shaping resilience, particularly from the perspective of complex adaptive systems. Thus, if social actors are fundamental to resilience, gaining a deeper understanding of their responses to unexpected shocks and events becomes critical, as does identifying the factors that influence their abilities and prospects in these scenarios. \cite{kurikka2021resilience} argued that deliberate and strategic human decisions can impact the trajectories of regional development, especially during crises. They also provided empirical evidence from two peripheral Finnish regions, underscoring that transformative agency is pivotal in fostering adaptability and reorientation post-shock.

The recent global crisis of COVID-19 has further intensified the focus on this perspective. COVID-19 has had a profound socio-economic impact, with the retail sector experiencing particularly unprecedented and widespread disruption. This turmoil is intricately connected to measures aimed at curbing the virus's spread and ensuring public safety, which have necessitated movement restrictions. Especially during the pandemic's initial stages, policies mandating the closure of non-essential retailers and service providers were implemented. Alongside concerns for personal safety, these measures led to significant and rapid changes in consumer spending and purchasing behaviors ~\citep{baker2020does, yang2021covid}. Therefore, it is without doubt that research is being vigorously pursued to comprehend the impact of COVID-19 countermeasures on individual and consumer behavior.

According to existing literature, consumers have increasingly favored shopping at local stores, smaller retail outlets, and locations outside of urban centers, shifting away from larger, more crowded urban retailers and city-center shops. In the U.S., mobile phone tracking data revealed a significant decline in foot traffic in areas with heavy commuter traffic, even before the implementation of official restrictions \citep{goolsbee2021fear}. In England, town centers saw a dramatic reduction in footfall, by as much as 75\%, with the extent of the decrease varying according to the size and type of the retail center. This indicates a marked decline in both transactions and consumption in these locales \citep{enoch2022covid}. Similar patterns emerged in Italy, where, during lockdowns, consumers showed a preference for `proximity' shops and online shopping over traditional supermarkets and street markets \citep{principato2022caring}.

One of the most dramatic and far-reaching changes in consumer behavior due to the pandemic was the widespread adoption of e-commerce, particularly during lockdowns, which significantly sped up the already growing online retail sector \citep{nanda2021would, oecd2020ecommerce} During the pandemic, not only did consumer shopping locations shift, but there were also notable changes in what people bought and how much they spent. Panic buying and stockpiling, triggered by economic uncertainty and occurring even before lockdowns, significantly altered spending patterns, particularly for essentials like food, alcohol, and hygiene products ~\citep{martin2020does, islam2021panic}

In summary, the previous research has primarily focused on examining what people bought, how often they shopped, and the location or platform they chose for the purchase, mainly qualitatively. However, it is essential to acknowledge the structural characteristics of consumer shopping trips or patronage behavior, considering the relationship between the product purchased and the connection between consumers' residence and shopping locations in terms of economic geography. For instance, the study of \cite{boyle2022impact}, which observed changes in consumer basket characteristics--the set of purchased products-- using store transaction data during the pandemic. The changes in consumer movement patterns and the composition of goods purchased due to the pandemic suggest that there have also been shifts in the patterns of shopping trips made by consumers to purchase a variety of amenity types. The shopping trips of consumers, which determine the scale and pattern of store agglomeration, are an essential feature in the literature related to location theory. Therefore, changes in these spatial consumption patterns should be considered when researching changes in consumer behavior. 

\subsection{consumer behavior and relatedness }
The principle of relatedness asserts that the growth rate of an activity within a location can be predicted by the presence of related activities. It is gaining prominence not only in academics but also plays a pivotal role in economic policies for regional development initiatives. This principle has been mentioned alongside other notable patterns in human geography~\citep{neffke2023evaluating}, including Zif's law for the distribution of city sizes ~\citep{zipf1946p}, the gravity law for social interactions ~\citep{Tinbergen1962}, Tobler's first law of spatial dependence ~\cite{tobler1970computer}, and the recognition of cities' exceptional role in the economy through urban wage and productivity ~\citep{rosenthal2004evidence, bettencourt2007growth}. 

Empirical research in the literature on the principle of relatedness commonly follows a three-step process~\citep{neffke2023evaluating}. Initially, the proximity between economic activities is established. Subsequently, for every activity within a region, the extent of related activities present is quantified. The final step involves analyzing how the growth rates of an activity are influenced by the presence of related activities in the region.  

To elaborate further, proximity can be estimated through the co-occurrence patterns of economic activities in a location. On this matter, \cite{hidalgo2021economic} describes as follows:

\begin{quote}
\ldots relatedness metrics can be used to estimate the combined presence of inputs that are specific to an activity, no matter if these inputs involve specific forms of labour, capital or institutions.
\end{quote}

Research in this area has typically been applied and studied in the context of the production of products and knowledge at the levels of technology, industry, and national units. When applied to knowledge production, this suggests that the effectiveness of regions in adopting a new economic activity is influenced not just by geographical and cultural closeness, like physical distance or shared language, but also by the cognitive and technological similarities between the new activity and the region's existing activities. 

Recently, \cite{hidalgo2020amenity} examined the co-location patterns of over one million amenities across 47 U.S. cities to analyze and optimize the mix of amenities in neighborhoods based on the proximity among amenities by extending the discussion of relatedness to the consumption sector. \cite{jun2022economic} utilized the analysis framework of \cite{hidalgo2020amenity} to explain the survival rates of amenity shops during the COVID-19 period through the relatedness metric, aiming to integrate the approach of relatedness with the discussion of resilience in the consumption sector. Their findings confirmed that a higher relatedness density in commercial areas increased survival rates during the pandemic. 

Despite the possibility of a relatedness approach to examine the structural characteristics of consumption patterns, research on this topic, aside from the two studies mentioned earlier, has been scarce. In particular, the studies by ~\cite{hidalgo2020amenity, jun2022economic} estimated and analyzed the relationship between amenities based solely on location data of stores not on actual consumption data. This paper aims to fill the gap in the existing literature by utilizing card data to calculate the consumption patterns of representative consumer groups aggregated from the data, using the relatedness approach. It also seeks to examine how these consumption patterns changed during the COVID-19 period. 

By applying the relatedness approach to consumption data, we estimate the complementarity between consumer purchases of amenities~\citep{luer2022leader, mendez2023store}, which influences consumer choices, and use this to explain the patterns of consumer shopping trips. \cite{clarke2006retail}, \cite{o1983impacts} and \cite{leszczyc2004effect} show that many customers engage in multipurpose, multi-stop trips. In other words, consumers perform multiple shopping activities along a single route in one location to save time and costs, which is directly related to the co-occurrence of consumption activities. This observation provides insight into the relationship between consumers' multi-purpose shopping trips and relatedness~\citep{ghosh1984model}. For a more detailed explanation, we will provide additional content in the following Section 3.4 on the economic model.

\section{Methods}
\subsection{Data}
To explore the consumers' behavior and the effect of relatedness density of amenity clusters on the behavior, we use two datasets: consumers' purchasing data compiled by BC card, a South Korean credit card company, and latitude and longitude store-location data compiled by the Korean Small Enterprise and Market Service. 

Our primary data source for analyzing consumption patterns is the transaction dataset provided by BC Card, spanning from 2018 to 2023. Due to limitations in data availability, our analysis is confined to the months of June and November of each year, specifically within Seoul. The dataset organizes transactions into a grid of 50m x 50m cells, covering 127 small-level and 9 large-level amenity categories: shopping, health/clinic and living-related services, food/drinks, retail/service for home and personal use, education, home and office repair services and supplies, entertainment/sports facilities, automotive services/facilities, accommodation. Additionally, it segments consumers based on age groups (e.g., 10s through 80s and above), life stages (e.g., single-person households, newlyweds with infants, elderly households, among others), and gender (female or male). The data also distinguishes between the locations of consumption and the residential areas of consumers, again using the 50m x 50m grid. Thus, it allows us to gain insight into the geographical distribution and volume of product purchases, aligned with detailed demographic and residential information of the consumers.

Our secondary dataset consists of store location data, including latitude and longitude, compiled by the Korean Small Enterprise and Market Service from 2018 to 2023. This dataset encompasses detailed information on small enterprises operating in Korea, classified into one of nine broad categories\footnote{travel/leisure/entertainment, real estate, retail, accommodation, sports facilities, restaurants, other living-related services, health, and education}. For each business, the dataset includes the store name, address, geo-location (latitude and longitude), and the types of amenities offered. We employ this dataset to determine the spatial unit of analysis, as discussed in Section~\ref{sc:spatialunit}. While this dataset is available from 2016 to 2023, we choose only the year 2018 so that we set the spatial unit of analysis based on the distribution of shops before the COVID-19 pandemic year, as well as considering the availability of the BC card transaction data. 

\subsection{Finding amenity clusters: the spatial unit of analysis}
\label{sc:spatialunit}
To investigate how changes in consumption behavior affect the resilience of amenity clusters during the COVID-19 pandemic, first, a meaningful spatial unit of analysis needs to be established. While administrative districts could be considered, they often fail to accurately reflect our daily consumption patterns due to a general lack of awareness of these administrative boundaries within urban environments. Consequently, we focus on delineating consumption spaces through the geographical locations of stores, providing a more accurate representation of urban consumption patterns beyond the administrative districts.

By using the location data of all small business shops in Seoul, we calculate the effective number of small businesses, $ A_{\alpha} $, and detect the amenity-dense neighborhood~\cite{hidalgo2020amenity, jun2022economic}. 
\begin{equation}
 A_\alpha = \sum^{N}_{\beta=1} e^{-\gamma d_{\alpha \beta}}
\end{equation} where $N$ denotes the total number of stores in the city and $d_{\alpha \beta}$ is the geodesic distance between stores $\alpha$ and $\beta$. From the standpoint of small business shops $\alpha$, the effect of the existence of other stores is diminishing over distance, and the decay parameter $\gamma$ captures this diminishing influence of distant stores. In our analysis, we set $\gamma$ into 7.58. This setting indicates that the influence of an amenity decreases by half every 91.44 meters and drops to 0.0022 at approximately 804.7 meters (about 0.5 miles), aligning with the median distance of daily walking trips, which typically take around 10 minutes~\citep{Yang2012}. Therefore, $A_\alpha$ represents a centrality score of the shop $\alpha$, which is the sum of the number of effective shops located within a 10-minute walking radius from the shop $\alpha$. 

Then, by looking at the spatial distribution of the effective number of shops, we identify the local peaks of $A_\alpha$ on the map. As we capture the center of the cluster with the local peak, we allocate neigh by shops to the cluster until the boundaries of the cluster overlap, resulting in 523 amenity clusters in Seoul, whose average radius is 241 meters. 

\subsection{Consumption space and relatedness density based on the multi-purpose trip of consumption}

Identified amenity clusters allow us to measure the distance/proximity among activities and build amenity space by looking at certain activities that happen in tandem within the clusters. For example, \cite{hidalgo2020amenity} and \cite{jun2022economic} construct the amenity space by looking at the co-location of certain types of shops, saying that the amenity space is a network connecting amenity types that are likely to co-locate in the same clusters. However, the collocation pattern of shops cannot directly represent the multi-purpose shopping behavior. Therefore, to capture the proximity among different amenity types based on consumption behavior, we look at the co-occurrence pattern of purchasing activities for consumer groups. Our data allows us to figure out where a consumer group purchases which types of products based on a 50m x 50m cell unit. Consumer groups are here defined as a combination of a specific residential area, age category, gender and a shopping area. This way, we can trace the consumer group's co-purchasing pattern in the detected amenity clusters, resulting in a total of 48,704 distinct consumer groups. In our study, to distinguish the amenity space constructed through co-purchasing patterns from the location-based amenity cluster, we refer to it as the ``consumption space''. 

Before calculating proximity, we capture only significant information of consumption behavior by using the revealed comparative advantage (RCA)~\citep{Balassa1965}. RCA of consumption of an amenity $p$ by the consumer group, $s$ in a cluster is the following:
\begin{equation}
    RCA_{s,p} = \left.{\frac{x_{s,p}}{\sum_{p}x_{s,p}}}\bigg/ \frac{\sum_{s}x_{s,p}}{\sum_{s,p}x_{s,p}}\right.
    \label{Eq:RCA}
\end{equation} where $x_{sp}$ is the consumption count of amenity $p$ of the consumer group $s$. 

By using the result of $RCA$, we measure the proximity between amenity $p$ and $p'$ following the way of \cite{Hidalgo2007}. The proximity $\phi_{p,p'}$ captures the conditional probability that two amenity types are co-purchased by the consumer groups in their single trip.
\begin{equation}
    \phi_{p,p'} = \left.min \Big\{P(RCA_{p}|RCA_{p'}), P(RCA_{p'}|RCA_{p}) \Big\}\right.
    \label{Eq:industrySimilarity}
 \end{equation} 

\begin{figure}[!t]
\centering
\includegraphics[width=1.0\linewidth]{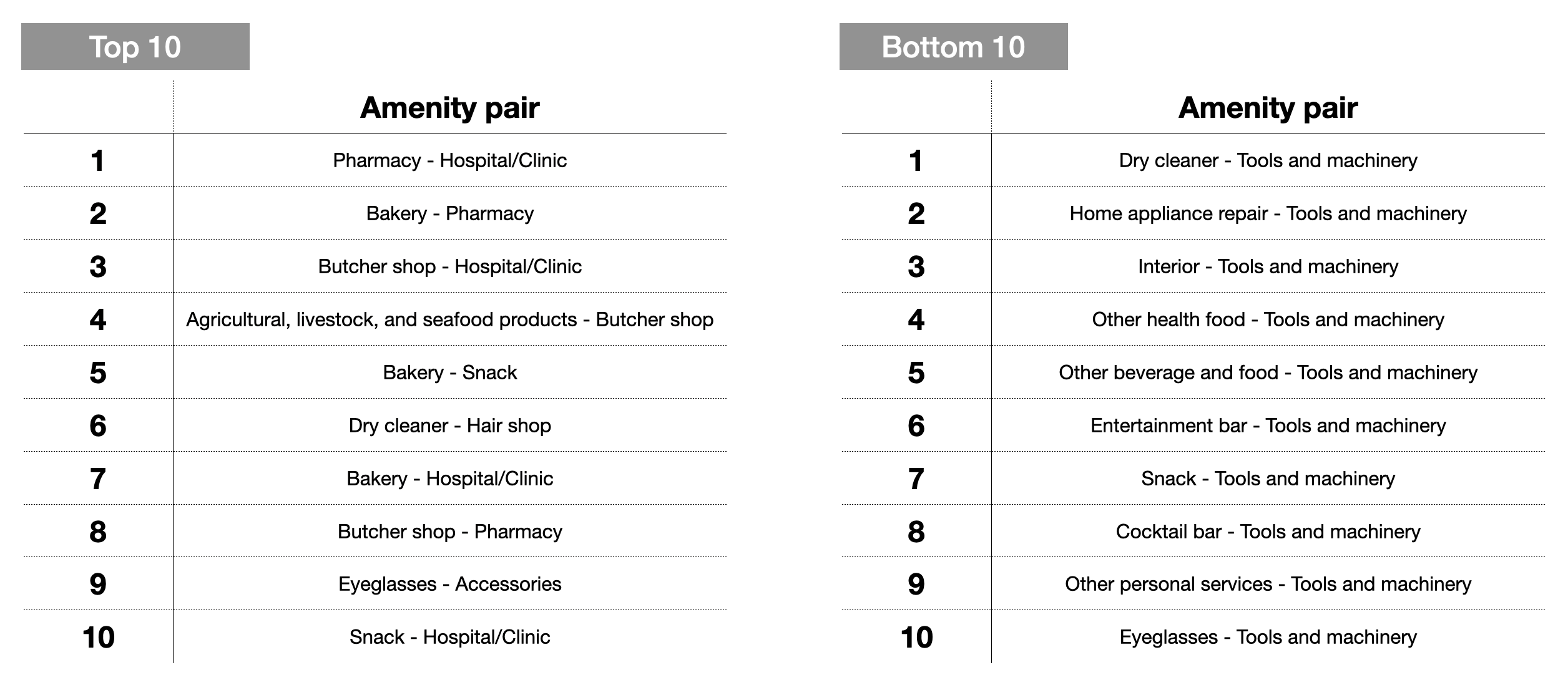}
\captionsetup{labelformat=empty}
\caption{\textbf{Table 1:} Top10 (left) and bottom10 (right) rankings of amenity pairs extracted based on calculated proximity.}
\label{pair}
\end{figure}
\setcounter{figure}{0}
\setcounter{table}{1}
Table~\ref{pair} shows the pairs of amenity types that have top 10 and bottom 10 proximity. The pharmacy and hospital/clinic are in the highest proximity. The highest proximity between the two reveals the fact that patients can purchase the prescribed medication only from the clinic in the pharmacy because of regulations in South Korea. The second and the third highest proximity also includes a Pharmacy and hospital/clinic. It might be because people who visit clinics are likely to visit bakeries and butcher shops in their single trip to the clinic since clinics are often located near residential areas. On the other hand, the bottom 10 list shows that machinery tools have very low complementarity with amenity types that are closely related to daily life. This implies that people who buy the machinery tools are likely to visit an amenity cluster with the aim of purchasing the specific product rather than making a multi-purpose trip.   

\begin{figure}[!t]
\centering
\includegraphics[width=1.0\linewidth]{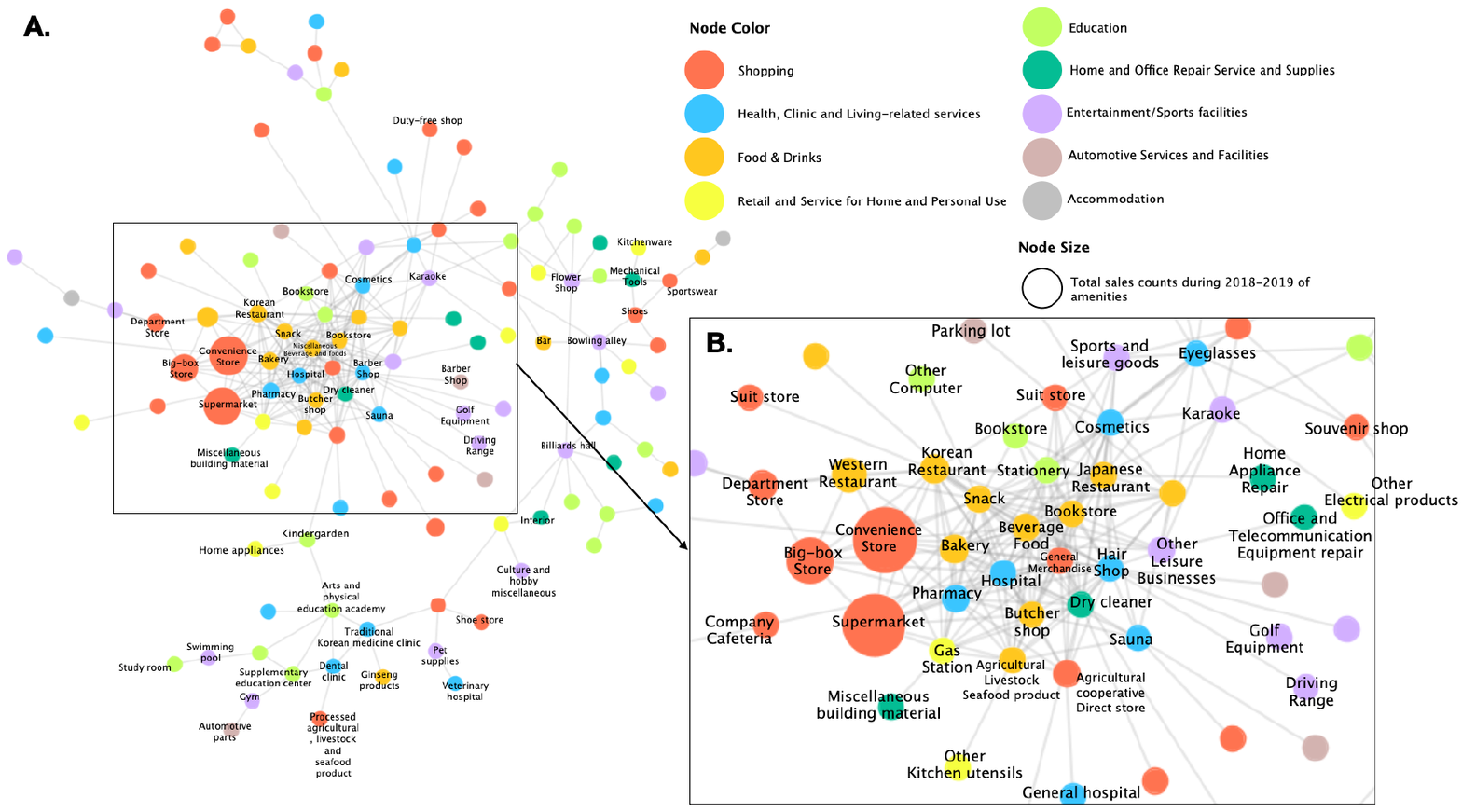}
\caption{Consumption space of residents in Seoul by looking at the co-purchasing behavior of consumer groups. Color represents types of amenities in large-level classification, while the size of the node represents the total counts for purchases during 2018-2019 in that amenity in Seoul: (A) the entire network, (B) A zoomed-in view of the network cluster}
\label{fig:consumptionSpace}
\end{figure}

Figure~\ref{fig:consumptionSpace} represents the network representation of proximity among amenities based on consumers' multi-purpose consumption behavior, i.e., consumption space, which reflects the shopping trip patterns of consumers purchasing several types of amenities in a single shopping area. Again, we can observe that the hospital/clinic and pharmacy are located in the center of the space, with surrounding amenities associated with our daily life, such as dry cleaners, butcher shops, supermarkets, and hair salons. We can observe that consumers are likely to co-purchase products of daily life in their shopping trips within one amenity cluster. 

At the bottom side of the figure, we can observe the network cluster, including various shops on private education and health care, such as supplementary education centers, gyms, swimming pools, study rooms, and arts and physical education academies. This cluster appears to reflect the prevalent private schooling among Korean students attending after-school academies in various disciplines such as music, math, writing, English, and martial arts. It can also reflect that parents or caregivers frequently accompany their children to these after-school programs and wait nearby while doing various activities. 

Based on the proximity among amenities calculated from consumers' multi-purpose trips, we construct the measure of relatedness $\omega_{i,p}$, where $i$ represents a shopping area and $p$ denotes an amenity type. The formula indicates a weighted average of the number of related amenities running in a shopping area based on consumers' co-purchasing behavior. Here, weights are the proximity between amenities $p$ and $p'$, $\phi_{p,p'}$ that we already calculated. Formally, the measure of relatedness density is given by:
\begin{equation}
    \omega_{i,p}= \frac{\sum_{p'}\phi_{p,p'}\cdot X_{i,p'}}{\phi_p} 
    \label{eq:density}
\end{equation} where $\phi_{p,p'}$ is the proximity between amenity $p$ and $p'$, $X_{i,p}=1$ if $RCA_{i,p}>1$\footnote{RCA of consumption of an amenity $p$ in a cluster $i$ is following:
\begin{equation}
    RCA_{i,p} = \left.{\frac{x_{i,p}}{\sum_{p}x_{i,p}}}\bigg/ \frac{\sum_{i}x_{i,p}}{\sum_{i,p}x_{i,p}}\right.
    \label{Eq:RCA} 
    \end{equation} where $x_{ip}$ is the consumption count of amenity $p$ in a cluster $i$. It is important to note that $RCA_{i,p}$ differs from $RCA_{s,p}$. When calculating the proximity between $p$ and $p'$, the objective is to capture the similarity in consumer groups' preference structures related to shopping trips. Here, the aim is to represent the extent of consumption activities within a specific cluster $i$, highlighting a distinction in purpose.} and otherwise. $\phi_p$ denotes $\sum_{p'}\phi_{p,p'}$. 

\subsection{Econometric model}
According to the literature on the principle of relatedness, countries, regions, and urban areas are more likely to enter a new activity, particularly in production sectors, when related activities are already established in them ~\citep{Hidalgo2007, Neffke2011,Boschma2015relatedness}. Although there has been growing empirical evidence on the principle of relatedness~\citep{hidalgo2018principle, hidalgo2021economic}, research on the consumption side is relatively under-explored. \cite{hidalgo2020amenity, jun2022economic} examined the amenity shops in the context of the principle of relatedness, focusing on the distribution of shops that consumers' choice can determine, but they cannot capture the consumption behavior itself. Moreover, when calculating relatedness density using consumption data, a different interpretation may be necessary compared to the density metric based on data related to knowledge, products, and technological production, which is on the production side. 

Here, we leverage consumption activity data to compute the relatedness density, denoted $\omega_{i,p,t}$. This metric measures the density of shopping activities for amenity types that are likely to be consumed together at the same location as amenity type $p$ at shopping area $i$ at time $t$. This essentially reflects the availability of multi-purpose shopping trips at location $i$ that include $p$. Consumers may choose to purchase or not a product $p$ based on their consumption preferences, when engaging in consumption activities of related types of amenities. Moreover, in our regression analysis, we add fixed effects for the (residential) origins of shoppers, the shopping district of destination, and amenities. Consequently, the coefficient of $\omega_{i,p,t}$ for purchase counts captures the extent of consumers' preference for multi-purpose shopping, controlling for the general shopping behavior for consumers from a given residential area, the overall attractiveness of a shopping area for any activity and the importance of a shopping activity across all locations. Therefore, we propose that if the coefficient of relatedness density is significant, it means that related shopping activities increase the likelihood of a purchase. The density coefficient can, therefore, be interpreted as an indication of \emph{consumers' preference for multi-purpose shopping trips}. 

In line with this, as consumers experienced the constraints of COVID-19, which restricted their shopping trips, what changes occurred in their behavioral patterns, particularly regarding their preferences for multi-purpose trips?  These trips, which involve purchasing activities for several types of amenities in a single journey, play a significant role in the development of urban commercial areas. Further, if there were changes, did they bounce back along with the recovery in the scale of consumption activities? or are they being sustained? To answer these questions, we construct the following empirical specification for three periods (pre-COVID: 2019, COVID: 2020-2022, Recovery\footnote{When examining the trend of purchase counts and amounts by year, we observed a declining trend starting from 2020, which rebounded in 2023. Therefore, we refer to 2023 as the ``Recovery'' year.}:2023) as follows:

\begin{equation}
    log(Y)_{i,j,p,t}= \beta_{0} + \beta_{1}\omega_{i,p,t} + \beta_{2}log(distance)_{i,j} + \beta_{3}\omega_{i,p,t} \cdot log(distance)_{i,j} + \mu_{i} + \eta_{j} + \gamma_{p} +\tau_{t} + \epsilon_{i,j,p,t}
    \label{Eq:interactioneq}   
    \end{equation} 
where $log(Y)_{i,j,p,t}$ represents the standardized log-transformed counts of purchases made by the consumer group residing in area $j$ for product $p$ in shopping area $i$ at time $t$. $\log(distance)_{i,j}$ denotes the log-transformed euclidean distance between shopping area $i$ and residence $j$. Along with our measure of relatedness $\omega_{i,p,t}$,  $log(distance)_{i,j}$ was also standardized. Variables standardized by adjusting the mean to 0 and the variance to 1 can be helpful in correctly analyzing interaction terms between $\omega_{i,p,t}$ and $log(distance)_{i,j}$.

In our model, we assume that consumers residing in $j$ have the option to visit any of the amenity clusters that we identify. As mentioned earlier, the consumer residence and shopping area employ the same spatial unit, implying that if consumers make purchases within the amenity cluster, including their residence, $i$ and $j$ become identical. The costs incurred by a shopping trip are symbolized by the shortest distance (i.e., the straight-line distance) between the consumer's residence and the shopping area~\citep {sevtsuk2021impact}. 

$\mu_{i}$ and $\eta_{j}$ are fixed effects for shopping and residence areas, respectively. They control for the time-invariant idiosyncratic characteristics of each spatial unit. When choosing a shopping location, consumers can also be influenced by the characteristics of commercial areas near their residences. Consequently, to capture the comprehensive effects of both shopping preference and residential influences, fixed effects extend beyond shopping locations to include residences. Additionally, we also adjust for the amenity fixed effect by employing its small-level amenity classification. In the pooled regression (as shown in columns (1) and (3) in Table \Ref{table:interaction}), we have also incorporated time-fixed effects to allow our results to control for national time trends. 

\section{Results}

\subsection{The effect of relatedness on consumption}
\label{section:main1}

\begin{table}[ht]
\centering
\caption{The effect of relatedness density with distance: all year(2019-2023), pre-COVID(2019), the COVID pandemic periods (2020-2023), and the recovery year (2023)}
\captionsetup[table]{font = {stretch = 1.5}, labelfont=it}
\label{table:interaction}
\begin{tabular}{lcccc}
   \tabularnewline \midrule \midrule
   Dependent Variable: & \multicolumn{4}{c}{$log(salescount)_{ijpt}$}\\
                        & All             & pre-Covid(2019) & Covid(2020-2022) & Recovery(2023) \\   
   Model:               & (1)             & (2)             & (3)              & (4)\\  
   \midrule
   \emph{Variables}\\
   $\omega_{ipt}$        & 0.0034          & 0.0029          & 0.0040           & 0.0018\\   
                        & (0.0033)        & (0.0047)        & (0.0036)         & (0.0031)\\   
   $\omega_{ipt}Xlog(distance)_{ij}$   & -0.1375$^{***}$ & -0.1427$^{***}$ & -0.1374$^{***}$  & -0.1335$^{***}$\\   
                        & (0.0021)        & (0.0038)        & (0.0020)         & (0.0024)\\
   $log(distance)_{ij}$     & -0.1536$^{***}$ & -0.1638$^{***}$ & -0.1506$^{***}$  & -0.1524$^{***}$\\   
                        & (0.0006)        & (0.0021)        & (0.0008)         & (0.0005)\\   
   \midrule
   \emph{Fixed-effects}\\
   Destination          & \checkmark      & \checkmark               & \checkmark                & \checkmark  \\  
   Residence            & \checkmark      & \checkmark               & \checkmark                & \checkmark  \\  
   Amenity              & \checkmark      & \checkmark               & \checkmark                & \checkmark  \\  
   Year                 & \checkmark      &                          & \checkmark                & \\  
   \midrule
   \emph{Fit statistics}\\
   Observations         & 166,461,440     & 33,292,288      & 99,876,864       & 33,292,288\\  
   $R^2$                & 0.04523         & 0.04455         & 0.04557          & 0.04585\\  
   Within $R^2$         & 0.04038         & 0.03877         & 0.04088          & 0.04114\\  
   \midrule \midrule
   \multicolumn{5}{l}{\emph{Clustered (Destination \& Residence) standard-errors in parentheses}}\\
   \multicolumn{5}{l}{\emph{Signif. Codes: ***: 0.01, **: 0.05, *: 0.1}}\\
\end{tabular}
\end{table}

Table~\ref{table:interaction} presents findings on how the relatedness density of an amenity cluster influences consumption patterns through multi-purpose trips. Column (1) of the table presents the result of pooled regression for all periods, and columns (2) to (4) show the results of each period, respectively. In all columns, $\omega_{ipt}$ is not statistically significant, whereas distance was found to be both negative and significant. The interaction term between $\omega_{ipt} \cdot log Distance_{ij}$ shows significantly negative effects on the counts of consumption that happen for amenity type $p$ in an amenity cluster $j$. The results suggest that the tendency for the purchase of the product with amenity type $p$ to increase at a given location is stronger when consumers make purchases close to their place of residence and as the level of relatedness of amenity $p$ within a cluster increases. According to literature, such as \cite{dellaert1998investigating}, consumers save on travel costs and time by visiting shopping areas with a diverse agglomeration of amenities for the multi-purpose shopping trip. In other words, our results imply that there may be a preference for multi-purpose shopping in areas close to residences and, therefore, shopping near residential areas that are accessible by walking explains the role of relatedness in consumer purchasing behavior more effectively than long-distance shopping trips do.

This consumption behavior, that consumers are likely to visit an amenity cluster with higher relatedness density when they visit nearby their house, appears to have weakened during the COVID-19 pandemic, and the weakening trend continued into 2023, despite a rebound in overall consumer spending during that year. However, it is difficult to check solely based on the depicted result in Table~\ref{table:interaction} that relatedness is effective in areas near consumers' residences since it merely implies that it works more effectively compared to long-distance trips. Therefore, we further analyze the effect of relatedness on consumption based on shopping distance, including any changes that occurred during the COVID-19 pandemic, by examining different distance intervals as shown in Section~\ref{section:distanceInterval}.

\subsection{Effect of relatedness of an amenity cluster on consumption over a trip distance of a consumer}
\label{section:distanceInterval}
If the effect of relatedness strengthens with shorter shopping distances, it prompts the question: Up to what distance does the density of relatedness show a significant effect? Does the impact of the COVID-19 pandemic vary with distance in terms of its effect on consumer purchasing behavior? To answer these questions, we first plot marginal effects of relatedness considering interaction terms with $log(distance)_{i,j}$ and also formalize this question by incorporating interaction terms for relatedness with COVID-19 and relatedness with recovery into \ref{Eq:interactioneq}, and by analyzing this equation across different distance intervals. The median daily walking trip distance is approximately 800 m (0.5 miles) to 1 km, with its walking duration being around 10 minutes ~\citep{Yang2012}. Moreover, the average daily distance traveled using public transportation by residents of Seoul is roughly 11 km, and the 90$th$ percentile for shopping distance is about 20 km. Based on this, we set the following seven distance intervals: 0 km, over 0 km to under 1 km, 1 km or more to under 2 km, 2 km or more to under 5 km, 5 km or more to under 10 km, 10 km or more to under 20 km, and over 20 km. The 0 km means that the residential location is within the same amenity cluster where shopping occurs, based on the spatial unit we used, indicating that the residence and the shopping cluster are identical.

\begin{figure}[!t]
\centering
\includegraphics[width=1.0\linewidth]{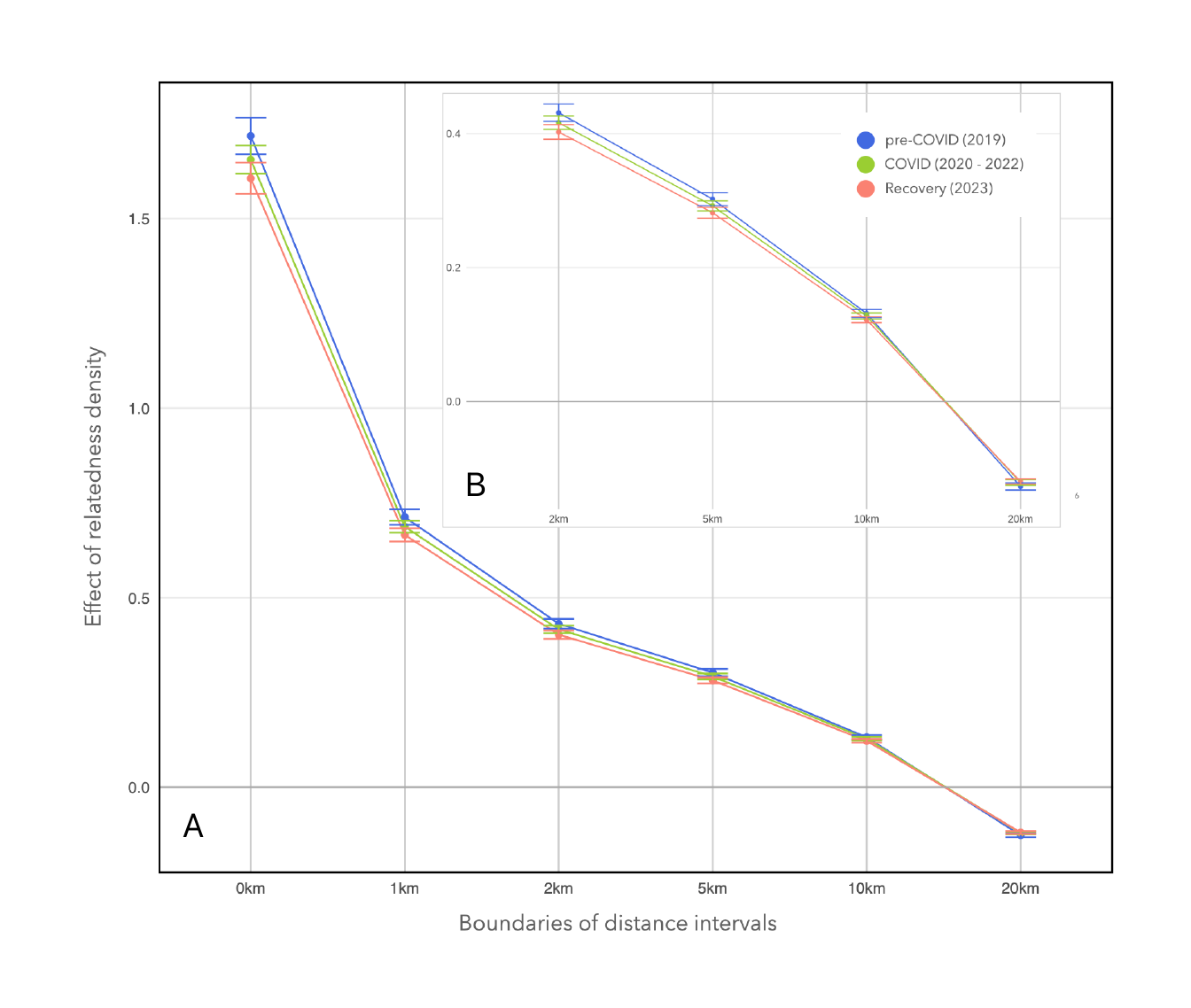}
\caption{Effect of relatedness density across boundaries of distance intervals: (A) covering all intervals, with $0km$ encompassing shopping trips within the commercial area adjacent to consumer's residence and $20km$ representing the 90$th$ percentile for shopping trip distances; (B) isolated plots for intervals beyond 2km.}
\label{fig:distomega}
\end{figure}

Figure \ref{fig:distomega} illustrates the effect of relatedness density across boundaries of distance intervals, plotted using the results from Equation \ref{Eq:interactioneq}. We calculated the effect of relatedness on $log(Y)_{i,j,p,t}$ by substituting standardized values of $log(distance)_{i,j}$ at the boundaries of distance intervals into the equation $\beta_{1}+\beta_{3}log(distance)_{i,j}$. The standard errors were calculated using the covariance between the two coefficients ($\beta_{1}$ and $\beta_{3}$ in equation \ref{Eq:interactioneq}). As shown, the effect of relatedness is positive at short distances and decreases as distance increases. Moreover, the downward trend during the COVID-19 pandemic (2020--2022) and recovery (2023) period is identifiable only within the short distance range (up to the 5km distance interval). 

To more clearly ascertain these results, we conduct regression analyses by distance interval as well. The sample is divided into seven groups based on the distance and the empirical specification is following:
\begin{equation}
    log(Y)_{i,j,p,t}= \beta_{0} + \beta_{1}\omega_{i,p,t} + \beta_{2}log(distance)_{i,j} + \beta_{3}\omega_{i,p,t} \cdot Covid + \beta_{4}\omega_{i,p,t} \cdot Recovery + \mu_{i} + \eta_{j} + \gamma_{p} +\tau_{t} + \epsilon_{i,j,p,t}
    \label{Eq:intervaleq}   
    \end{equation} where $Covid$ is a dummy variable that takes the value 1 for the years 2020 to 2022, reflecting the period of the COVID-19 pandemic when there was a general trend of decline in consumption quantities and amounts due to restricted mobility and related economic challenges, and otherwise 0. The $Recovery$ dummy variable, on the other hand, takes the value 1 for the year 2023, indicating a period where the trend of decline in consumption is expected to rebound, potentially due to easing restrictions, economic stimulus, and adaptation to the pandemic's conditions. The base for the COVID and recovery dummy variables is the year 2019. 

Table \ref{table:dinterval} shows the results that estimate of whether the relatedness density influences consumer purchase choices and how preferences for multi-purpose shopping trips vary across different distance intervals. The significant positive relationship between relatedness density and purchase count implies that the consumption of amenity types, which are more likely to be consumed in the same place, facilitates each other's purchase decisions. Therefore, we previously mentioned this as an explanation for the preference towards multi-purpose shopping trips. 

Column (1) of Table \ref{table:dinterval} corresponds to the result considering purchases with a 0 km shopping distance, which indicates that the consumer's place of residence and that of purchase are identical within our spatial unit. First, we find the positive effect of $\omega_{i,p,t}$ on consumption decisions, confirming the preference for multi-purpose shopping trips in close proximity to the consumer's residence. This supports the findings of the study by ~\cite{leszczyc2004effect}, which examined the shopping strategies of different consumer types and found that consumers who utilize multi-purpose shopping trips as a strategy - identified as time-constrained service seekers and time-constrained price seekers - prefer shopping in area nearby residence with higher-level of relatedness density due to their emphasis on efficiency. Moreover, results in Column (1) exhibit a relatively high $R^2$ (0.68682) for purchase decisions made in shopping areas near the consumer's residence. We also discover that while the magnitude and significance of the $\omega_{i,p,t}$ coefficient, as well as the $R^2$ value, decrease as distance increases, the relatedness density remains positive and significant up to the 5km distance interval. These findings are consistent with what was observed in Figure \ref{fig:distomega}. 

The results for the COVID-19 and recovery dummy variables are quite interesting. While Figure~\ref{fig:distomega} shows a continuous weakening of the marginal effect of relatedness density across all intervals up to 2023, columns (1) - (3) indicate that although the trend of weakening is consistent with negative coefficients for COVID and Recovery dummy, the reduction in the effect during the COVID-19 period is more significant than during the recovery period when compared to 2019. This finding does not suggest a continuous weakening but rather indicates that following a decline during the COVID-19 pandemic, there is a rebound in the preference for multi-purpose shopping trips within a 2km distance, along with a recovery in the scale of consumption. In other words, the result implies that consumer behavior patterns for shopping trips within 2km are resilient regarding multi-purpose shopping trips. 

Conversely, columns (4) and (5), which present the results for shopping trips within distances of more than 2km but less than 10 km, show that the trend of weakening in the effect of relatedness density observed during the COVID-19 period continues until 2023. 

\begin{landscape}
\begin{table}[ht]
\centering
\caption{Distance interval analysis: Changes during the COVID period (2020-2022) and recovery period is (2023) compared to the base year 2019}
\label{table:dinterval}
\resizebox{\linewidth}{!}{%
\begin{tabular}{lccccccc}
   \tabularnewline \midrule \midrule
   Dependent Variable: & \multicolumn{7}{c}{$log(salescount)_{ijpt}$}\\
                      & $distance = 0 $           & $0<distance\leq1km $            & $1km<distance\leq2km  $                   & $2km<distance\leq5km $            & $5km<distance\leq10km$                    & $10km<distance\leq20km$                   &$ distance>20km$ \\   
   Model:             & (1)             & (2)                 & (3)                  & (4)                  & (5)                   & (6)                    & (7)\\  
   \midrule
   \emph{Variables}\\
   $\omega_{ipt}$      & 1.256$^{***}$   & 0.3375$^{***}$      & 0.0940$^{***}$       & 0.0078$^{*}$         & 0.0017                & 0.0017                 & 0.0011\\   
                      & (0.0785)        & (0.0389)            & (0.0210)             & (0.0040)             & (0.0036)              & (0.0031)               & (0.0008)\\   
   $\omega_{ipt}XCovid $    & -0.2837$^{***}$ & -0.1236$^{***}$     & -0.0708$^{***}$      & -0.0103$^{***}$      & -0.0035$^{***}$       & -0.0017$^{**}$         & -0.0006$^{*}$\\   
                      & (0.0450)        & (0.0207)            & (0.0134)             & (0.0018)             & (0.0011)              & (0.0009)               & (0.0003)\\   
   $\omega_{ipt}XRecovery$  & -0.1980$^{***}$ & -0.1020$^{***}$     & -0.0683$^{***}$      & -0.0114$^{***}$      & -0.0040$^{**}$        & -0.0030                & -0.0010\\   
                      & (0.0522)        & (0.0256)            & (0.0141)             & (0.0025)             & (0.0020)              & (0.0024)               & (0.0007)\\   
   $log(distance_{ij})$   &                 & -0.8610$^{***}$     & -0.6490$^{***}$      & -0.0352$^{***}$      & -0.0048$^{***}$       & -0.0011$^{***}$        & 0.0009$^{*}$\\   
                      &                 & (0.0808)            & (0.0413)             & (0.0038)             & (0.0010)              & (0.0003)               & (0.0005)\\   
   \midrule
   \emph{Fixed-effects}\\
   Destination                 & \checkmark              & \checkmark                  & \checkmark                   & \checkmark                   & \checkmark                    & \checkmark                     & \checkmark \\  
   Residence              & \checkmark              & \checkmark                  & \checkmark                   & \checkmark                   & \checkmark                    & \checkmark                     & \checkmark \\  
   Amenity         & \checkmark              & \checkmark                  & \checkmark                   & \checkmark                   & \checkmark                    & \checkmark                     & \checkmark \\  
   Year               & \checkmark              & \checkmark                  & \checkmark                   & \checkmark                   & \checkmark                    & \checkmark                     & \checkmark \\  
   \midrule
   \emph{Fit statistics}\\
   Observations       & 325,120         & 1,035,050           & 3,224,530            & 17,717,770           & 46,032,420            & 78,798,420             & 19,328,130\\  
   R$^2$              & 0.68682         & 0.28326             & 0.07898              & 0.01277              & 0.00703               & 0.00760                & 0.00583\\  
   Within R$^2$       & 0.00772         & 0.00278             & 0.00364              & 0.00026              & $2.23\times 10^{-5}$  & $8\times 10^{-6}$      & $9.93\times 10^{-6}$\\   
   \midrule \midrule
   \multicolumn{8}{l}{\emph{Clustered (Destination \& Residence) standard-errors in parentheses}}\\
   \multicolumn{8}{l}{\emph{Signif. Codes: ***: 0.01, **: 0.05, *: 0.1}}\\
\end{tabular}%
}
\end{table}
\end{landscape}

As previously mentioned, distance represents the cost incurred by consumers when deciding on a shopping trip. Therefore, it is common for the number of consumer purchase decisions to decrease as distance increases. For estimators of $\log(distance)_{i,j}$ in Table \ref{table:dinterval}, the value decreases as distance increases across columns (2) to (7), indicating that consumer purchase decisions are more sensitive to the distance at shorter shopping distances. Especially once the distance exceeds 2km, the elasticity of consumer purchase behavior in response to distance severely decreases, suggesting that there may be a fundamental difference in consumer behavior between shopping trips under 2km and those that involve traveling more than 2km. 

\subsection{The relationship between the type of amenity clusters and consumers' consumption behavior}
Which characteristics do clusters that consumers travel long distances to shop in possess? What do consumers purchase through long-distance trips? To enhance our understanding of consumers' preference for multi-purpose shopping near their residences and to gain insights into long-distance shopping trips, we grouped types of amenity clusters using the k-means clustering method, organizing them based on population characteristics -- residential population density, floating population density, and working population density. 

\begin{figure}[!t]
\centering
\includegraphics[width=1.0\linewidth]{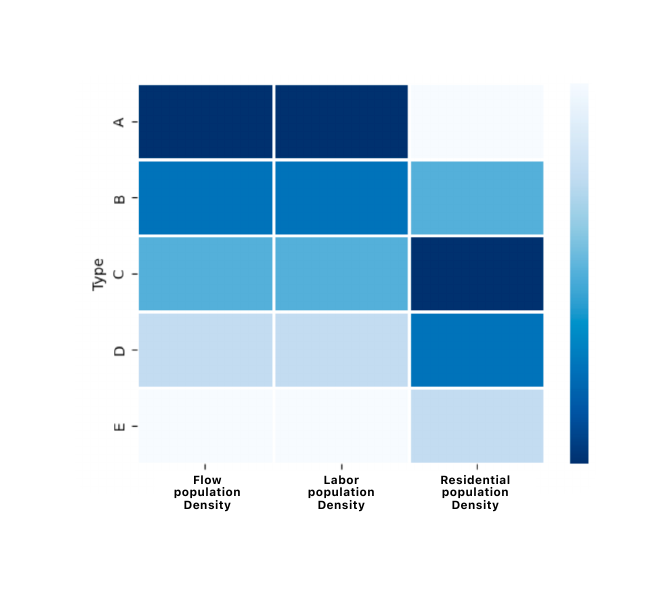}
\caption{The ranking of floating population density, working population density, and residential population density by type of amenity cluster (with denser areas indicated by darker colors: we perform k-means clustering on the vector for each amenity cluster [Flow population density, Labor population density, Residential population density] to categorize types of clusters.)}
\label{fig:type}
\end{figure}

Figure ~\ref{fig:type} illustrates the characteristics of the floating population, working population, and residential population by cluster type. As shown, type A clusters are characterized by the highest densities of floating and working populations, encompassing commercial areas where industrial complexes or large shopping malls are located. A notable feature is the lowest residential population, indicating these are highly concentrated commercial districts. Type B clusters are downtown commercial areas with the second-highest densities of floating and working populations and the third-highest residential population density. These are frequently visited by tourists as well as residents of Seoul, making them vibrant hubs of economic activity. Type C and D can be described as residential commercial areas, ranking first and second, respectively, in residential population density. Type E includes areas with green spaces or commercial areas located on the outskirts. 

We split the sample over the cluster types of places where consumption happens, using the same empirical specification, Equation~\ref{Eq:intervaleq}. Table \ref{table:type} shows the results with the split sample. Firstly, in line with previous findings, cluster types C and D, which have high residential population densities, relatedness density is found to be positive and highly significant. This result confirms our previous findings that relatedness density matters in consumers' decisions with a clear preference for shopping at amenities close to consumers' residences. Type A, characterized by concentrated specific industries such as industrial complexes or locations with large shopping malls, appears to be mainly active in long-distance shopping activities. Similarly, while Type B has a relatively substantial residential population, as a downtown commercial district, the prevalence of consumption through long-distance trips suggests that the preference for multi-purpose shopping trips, as explained by relatedness density, may not be observed as can be seen in the Table \ref{table:type}. 

The results on distance elasticity further support previous findings. In commercial areas like Type A and B, which have a high proportion of long-distance trips, the reduction in consumer activities as distance increases is relatively smaller compared to other types. Meanwhile, types C and D, primarily characterized by higher residential population densities, show similar estimates.

\begin{table}[ht]
\centering
\caption{Effect of relatedness density across the different types of amenity clusters that the consumption happen: Type A indicates industrial complex or locations with large shopping malls; Type B clusters are downtown commercial areas; Type C and D are characterized by residential and commercial areas; Type E includes areas with green spaces or commercial areas located on the outskirts. }
\captionsetup[table]{font = {stretch = 1.5}, labelfont=it}
\label{table:type}
\begin{tabular}{lccccc}
   \tabularnewline \midrule \midrule
   Dependent Variable: & \multicolumn{5}{c}{$log(salescount)_{ijpt}$}\\
                      & type A          & typeB           & typeC           & typeD           & typeE \\   
   Model:             & (1)             & (2)             & (3)             & (4)             & (5)\\  
   \midrule
   \emph{Variables}\\
   $\omega_{ipt}$      & 0.0013          & 0.0161          & 0.0038$^{**}$   & 0.0058$^{***}$  & 0.0032$^{*}$\\   
                      & (0.0047)        & (0.0154)        & (0.0015)        & (0.0012)        & (0.0017)\\     
   $\omega_{ipt}XCovid $    & -0.0055         & -0.0144$^{***}$ & -0.0041$^{***}$ & -0.0035$^{***}$ & 0.0002\\   
                      & (0.0043)        & (0.0033)        & (0.0011)        & (0.0008)        & (0.0010)\\   
   $\omega_{ipt}XRecovery$  & -0.0031         & -0.0197$^{**}$  & -0.0035$^{***}$ & -0.0032$^{***}$ & 0.0020\\   
                      & (0.0042)        & (0.0090)        & (0.0010)        & (0.0009)        & (0.0015)\\   
    $log(distnace)_{ij}$   & -0.0593$^{***}$ & -0.1137$^{***}$ & -0.1701$^{***}$ & -0.1701$^{***}$ & -0.1607$^{***}$\\   
                      & (0.0090)        & (0.0074)        & (0.0041)        & (0.0043)        & (0.0141)\\  
   \midrule
   \emph{Fixed-effects}\\
   Destination                 & \checkmark              & \checkmark              & \checkmark              & \checkmark              & \checkmark \\  
   Residence             & \checkmark              & \checkmark              & \checkmark              & \checkmark              & \checkmark \\  
   Amenity         & \checkmark              & \checkmark              & \checkmark              & \checkmark              & \checkmark \\  
   Year               & \checkmark              & \checkmark              & \checkmark              & \checkmark              & \checkmark \\  
   \midrule
   \emph{Fit statistics}\\
   Observations       & 5,527,040       & 31,211,520      & 61,772,800      & 56,570,880      & 11,379,200\\  
   R$^2$              & 0.01397         & 0.02070         & 0.03001         & 0.03241         & 0.03251\\  
   Within R$^2$       & 0.00484         & 0.00867         & 0.02473         & 0.02814         & 0.02733\\  
   \midrule \midrule
   \multicolumn{6}{l}{\emph{Clustered (Destination \& Residence) standard-errors in parentheses}}\\
   \multicolumn{6}{l}{\emph{Signif. Codes: ***: 0.01, **: 0.05, *: 0.1}}\\
\end{tabular}
\end{table}

Throughout the COVID-19 pandemic, commercial area types C and D show a trend where the effect of relatedness density decreases during the COVID-19 period and then bounces back during the recovery period. Type E, which does not show significant interactions with COVID and recovery variables, appears unaffected by the pandemic, likely due to its location on the outskirts where overall population density is lower. In contrast, the downtown commercial areas, Type B, show a continued effect of weakened relatedness density from the COVID period into the recovery period. We expect that Type A would show a similar trend to Type B, but it does not show significant results for the two dummy variables. The reason for the distinct pattern exhibited by Type A can be understood through the network of shopping trips between commercial areas and more explanation on it is in Appendix B.\footnote{In Appendix B. we visualized a network of consumer's trip. According to Figure~\ref{fig:clusterspace} in Appendix B, shopping trips to Type A are almost sparse compared to Type B. Given that this study is limited to shopping trips within Seoul by residents of Seoul. It appears that consumption area Type A, which includes industrial complexes and large shopping malls that also attract visitors from outside Seoul, has not been considered.}


\section{Conclusion and discussion}
Recognizing the significance of human activities in shaping the resilience of regional economies is certainly one issue, but there's a growing acknowledgment of the need for a fundamental understanding of how social agents respond and adapt to economic shocks in complex regional economies~\citep{bristow2014regional, kurikka2021resilience}. In line with this literature stream, this study explores the influence of relatedness density within amenity clusters on consumer purchasing behavior, with a specific focus on the preference for multi-purpose shopping trips. It analyzes how this relationship changes over various shopping distances and in response to the COVID-19 pandemic. 

Our findings demonstrate that relatedness has a positive and significant effect on consumer purchases at shorter distances. This suggests a preference for multi-purpose shopping trips near consumers' residences. This effect diminishes with increasing distances, with the impact of relatedness becoming less significant beyond 2km. During the COVID-19 pandemic, a weakening effect of relatedness density on consumption is observed, especially within shorter distances. However, this trend shows signs of recovery in 2023, indicating a resilience in consumer behavior patterns towards multi-purpose shopping trips within 2km from their residences. This implies that despite the pandemic's impact, there remains a consistent preference for shopping close to home when it involves amenities related in type. 

Furthermore, we classified amenity clusters into five types based on population characteristics (floating, labor, and residential population density) and analyzed the impact of relatedness density on consumption within these clusters. We found that clusters characterized by residential and commercial areas (Types C and D) exhibit a significant positive relationship between relatedness density and purchase count, aligning with the overall preference for shopping in proximity to residential locations. Conversely, areas characterized by long-distance shopping trips (Types A and B) show varied responses to relatedness density and distance elasticity. A downtown commercial district (Type B) experienced a continued weakening in the effect of relatedness density from the COVID-19 period into the recovery phase, while industrial complexes or locations with large shopping malls (Type A) appear unaffected by the pandemic in terms of multi-purpose shopping behavior. 

\begin{figure}[!t]
\centering
\includegraphics[width=1.0\linewidth]{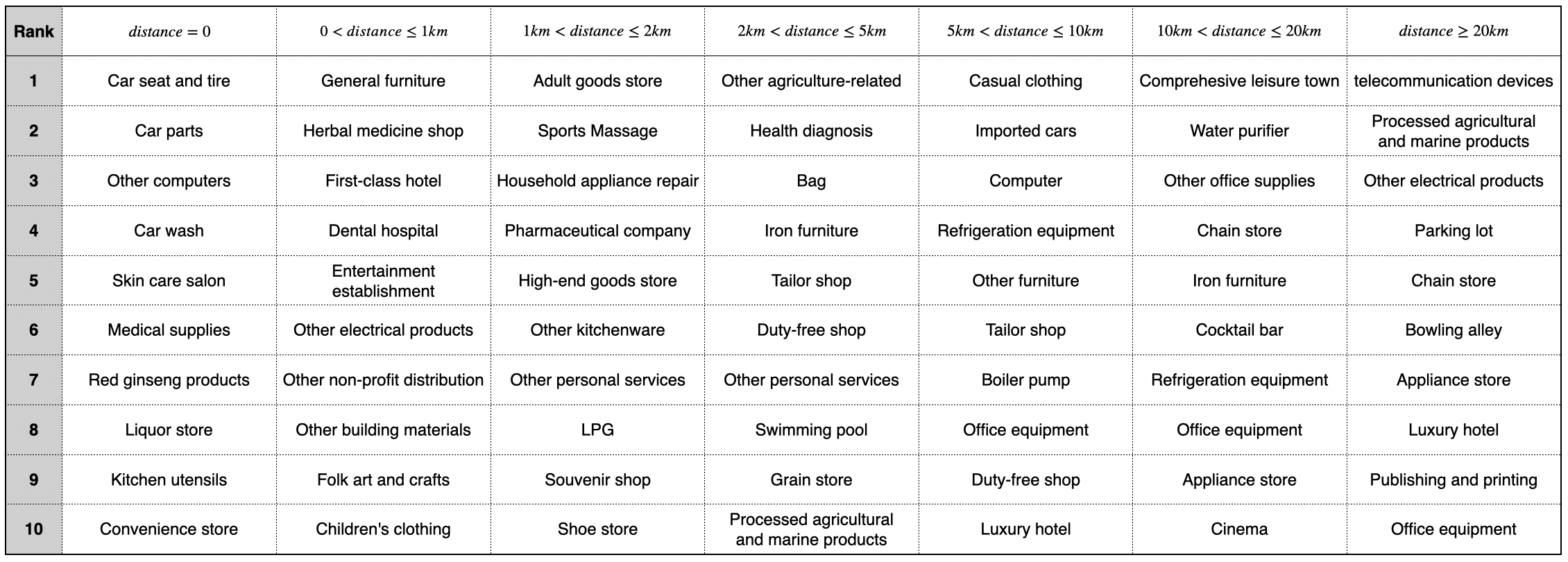}
\caption{Top 10 amenity types over distance}
\label{fig:rank}
\end{figure}

Through our results, we confirmed that clear differences exist in consumer shopping behaviors based on distance. Then, what differences exist in consumption patterns according to distance? To answer the question, we have identified the types of amenities that are relatively more frequently purchased in each distance interval, emphasizing the relative, not absolute, magnitude of purchases. This is calculated using the following formula to ascertain which amenities are purchased more in comparison to other intervals:

\begin{equation}
    X_{d,p} = \left.{\frac{x_{d,p}}{\sum_{p}x_{d,p}}}\bigg/ \frac{\sum_{d}x_{d,p}}{\sum_{d,p}x_{d,p}}\right.
    \label{Eq:rank}
\end{equation} where, $X_{d,p}$ represents the consumption count of amenity $p$ within distance interval $d$. This approach allows for the measurement of structural characteristics of consumption patterns across distance intervals, offering a nuanced understanding of how consumer preferences shift in relation to distance from amenities. Figure ~\ref{fig:rank} displays the top 10 amenity types according to $X_{d,p}$, highlighting the predominant amenities preferred relatively by consumers within the specified distance interval. As shown, at closer distances, amenity types closely related to daily life, such as car parts, car wash, and skin care, while at further distances, amenity types that are less frequently consumed, are included, such as telecommunication devices, electronic devices, appliance stores, and water purifiers stand out. 

To understand our observation, the concept of comparison shopping is required together with multi-purpose shopping. While multi-purpose shopping establishes a theoretical foundation for the spatial agglomeration of shops across various sectors~\citep{hanson1980spatial, o1981model}, comparison shopping describes the patterns where consumers, due to a lack of information about product quality or price, visit multiple stores of the same type to gather information and compare options~\cite{wolinsky1983retail, brueckner2011lectures}. Electronic and household appliances are typical items for consumers to compare when shopping. Consumers visit clusters of similar businesses, even if they are somewhat distant, to engage in comparison shopping for such products. In particular, consumers who consider not only price but also personal preferences, called cherry pickers, tend to visit commercial districts where similar stores are clustered to facilitate product exploration. For these consumers, the importance of the distance to those districts is less significant compared to other groups~\citep{leszczyc2004effect}. Therefore, by integrating the results of previous analyses with the ranking of amenity types by distance, we suggest that the impact of relatedness and the elasticity in relation to distance appear differently because short distances are associated with multi-purpose shopping trips, while the behavioral pattern of comparison shopping or shopping by cherry picker types facilitate longer distance shopping trips.

As ~\cite{Boschma2015resilience} has highlighted, developing the concept of regional economic resilience from an evolutionary perspective requires integrating both short- and long-term perspectives. This may apply equally to understanding the resilience of consumer behavior. Although we examine the effect of relatedness density on consumer behavior and how COVID-19 influences the effect, our study is limited in observing the long-term effects of such shocks. Due to issues related to data availability and the fact that signs of recovery following the COVID-19 pandemic were primarily observed in 2023, our research is constrained to focusing on the short-term relationship between relatedness and consumer behavior. 

Limited categories and classifications of amenities restrict discussions on the micro-mechanisms of relatedness density that explain consumer behavior as well. \cite{neffke2023evaluating} highlights that essential aspects of the principle of relatedness are not well understood, although the principle of relatedness, which proposes notable empirical regularities, plays a crucial role not only in academic work but also in economic policy frameworks underpinning large-scale place-based development programs. This study extends the discussion on relatedness to the consumption side, presenting and empirically demonstrating the association with consumers' shopping behaviors, such as multi-purpose trips. However, a more detailed understanding requires further discussion and validation. For example, department and big-box stores host a variety of businesses, enabling consumers to engage in multiple shopping activities internally, yet the data we utilized does not differentiate classifications of amenity for these activities. In addition, the most significant change in consumer behavior during the COVID-19 pandemic can be said to be the increase in online consumption. However, this study focuses solely on offline shopping. 

Despite our limitations, our results shed light on our understanding of regional economic resilience in terms of consumer behavior, which will be the subject of future research. While our study provides insightful observations on the resilience of consumer behavior in regional economies, focusing exclusively on offline shopping trips, it's important to recognize a significant shift in consumer behavior towards online shopping, prompted by the COVID-19 pandemic. This shift to online platforms presents an opportunity to investigate how the expansion of online shopping influences offline shopping behaviors and consumer preferences. Moreover, changes in consumer behavior are not only driven by the shift between offline and online shopping but also by basic demographic characteristics such as age and gender, as well as household composition trends, such as the increase in single-person and elderly households. Future research could explore how these demographic and household characteristics influence consumption patterns, particularly in response to economic shocks like the pandemic. By considering these aspects, studies could provide a more comprehensive understanding of how consumer preferences and behaviors evolve in the face of significant societal shifts, thereby contributing to the broader discourse on economic resilience and consumer dynamics in both online and offline shopping contexts.

\section*{Acknowledgement}
This project is funded by the National Research Foundation of Korea (NRF-2022R1A5A7033499). We also acknowledge the support of the Inha University. 


\linespread{1.5}

\newpage
\linespread{1.0}
\appendix
\begin{appendices}
\setcounter{table}{0}
\setcounter{figure}{0}

\newpage
\numberwithin{equation}{section}
\section*{Appendix A. Network Representation for Consumers' Shopping Trips}
The network($o-d$ network) between consumers' residential areas(origins, $o$) and shopping amenity clusters(destinations, $d$) is useful for visually verifying the consumers' shopping trips. Creating a network for consumers' shopping trips in Seoul involves mapping out the interactions between thier residences and destinations (shopping areas) within the city. We have constructed "$o-d$ network" for consumers' shopping trips across three periods: pre-COVID (2019), during COVID (2020-2022), the onset of recovery phase(2023), using card transaction data. 

By examining the changes in the network across three periods, we aim to analyze how consumer shopping trip patterns have qualitatively and structurally evolved through the COVID pandemic. Specifically, where do consumers move from and to for the their shopping needs, and how has the pandemic transformed these shopping patterns? Does the quantitative recovery observed in 2023 signify a return to pre-pandemic patterns, or have new shopping behaviors emerged? Through the analysis of the shopping trip network, we seek to answer these questions. By utilizing card transaction data, we capture a comprehensive picture of shopping activities, providing insights into both the resilience and adaptability of consumer habits in the face of global disruption like the COVID pandemic. 

\vspace{\baselineskip}
\subsection*{The characteristics of Nodes}
According to our results, it appears that short distance shopping near residential areas and long distance shopping are based on different underlying mechanisms. Therefore, short distances shopping is represented by nodes, while long distance shopping is represented through edge. In this context, we define the characteristics of a node as follows: firstly, the size of nodes is calculated by \ref{Eq:nodesize}

\begin{equation}
    X_{d\leq1km,(i,j),t} = \sum_{p}x_{d\leq1km,(i,j),p,t}
    \label{Eq:nodesize}
    \end{equation} where $x_{d\leq1km,(i,j),p,t}$ denotes the number of purchases made for the type of amenity, $p$, by consumers residing in area $j$ and shopping in  area $i$ at time $t$. $d$ represents the distance for shopping trips from $j$ to $i$. 

The color of each node has been assigned based on the categorization of amenity clusters, which as determined using k-means clustering. This classification is rooted in three key demographic densities as discussed in the earlier sectors of this paper, floating population density, labor population density, and residential population density. Through the k-means clustering approach, we have divided the amenity clusters into five distinct categories, labeled A to E. Each category reflects a unique combination of these demographic densities, which in turn signifies the potential consumer reach and the nature of economic activities prevalent in each cluster. This classification allows for a nuanced visualization of the network, facilitating a deeper understanding of the spatial dynamics and economic potentials of different areas within the city. 

\vspace{\baselineskip}
\subsection*{The characteristics of Edge}
 Following the explanation of nodes within $o-d$ network, Edges represent long-distance shopping trips, providing insight into the broader movement patterns that extend beyond shopping trips nearby consumers' residence. This typically encompasses shopping trips that involve traveling by public transportation or car, rather than those made on foot. Each edge in the network is directed, indicating the flow from a consumers' residence (the source) to a shopping destination (the target). 
 Therefore, the edge's weight is determined by the total number of purchases made during trips defined by a specific source and target pair. This is presented as:
 
\begin{equation}
    X_{d>1km,(i,j),t} = \sum_{p}x_{d>1km,(i,j),p,t}    
    \label{Eq:edge}
    \end{equation} where $x_{d\leq1km,(i,j),p,t}$ denotes the same in equation \ref{Eq:nodesize}.

The color of the edge is assigned based on the cluster type of the target. In other words, the cluster type of the destination determines the edge color. This approach allows us to identify not just the frequency of trips between specific areas but also he intensity of shopping activities, as evidenced by purchase counts. 

\vspace{\baselineskip}
\subsection*{The transformation of Shopping trips Network throughout the COVID-19}
Using a network representation for the $o-d$ network, we can identify pairs of amenity clusters experience high purchase frequencies, discern the groups they form, and understand their types of clusters as well as the extent of local shopping activity. This approach not only reveals a rough picture of the most frequented shopping trip corridors but also categorizes them into discernible the type of clusters, highlighting the varying levels of shopping intensity across the city.

However, it's crucial to acknowledge that while network representation serves as a powerful visualization technique, it is not an end in itself. Instead, it provides a means to visually verify and complement the descriptive characteristics of shopping behaviors. As such, the network visualization aids in the preliminary understanding of spatial and transactional patterns, but is alone cannot capture the full complexity of consumer behavior or the underlying reasons for these patterns. Therefore, this network analysis is employed primarily as a supplementary tool to enhance our main findings. 

\begin{figure}[!t]
\centering
\includegraphics[width=1.0\linewidth]{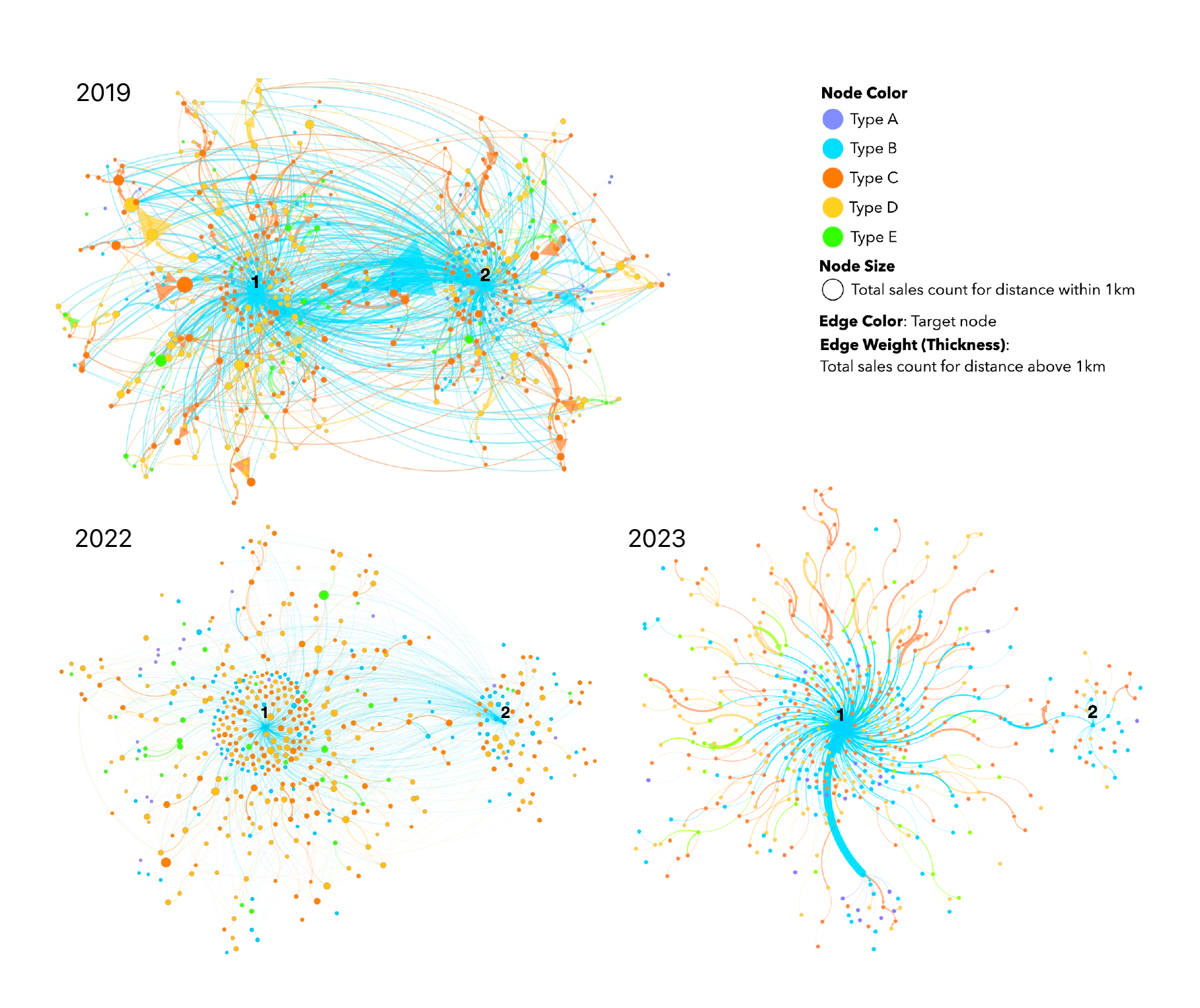}
\caption{}
\label{fig:clusterspace}
\end{figure}

The characteristics of each cluster type can be described as follows: type A clusters stand out due to their exceptionally high densities of floating (people moving through the area) and labor (working) population as shown in Firgure ~\ref{fig:type}, primarily found in zones housing industrial complexes or large shopping malls. These areas are marked by the lowest residential density, highlighting their role specialized commercial districts rather than living spaces. Conversely, type B clusters represent thriving downtown commercial clusters with notable levels of both floating and labor population, along with a moderate residential density. This combination makes them attractive to both tourists and local Seoul residents, serving as key economic centers. Types C and D are characterized as residential commercial areas with, Type C leading in residential population density followed closely by Type D. Lastly, Type E encompasses areas with abundant green spaces or those located on the city's periphery.

We predicted that commercial areas classified as type A, known for their high pedestrian and labor population density, represent as the primary destinations for long-distance shopping trips. However, two amenity clusters in type B came first as central hubs throughout all periods. The first location is a downtown commercial area (No. 1 in Figure B\ref{fig:clusterspace}, Type B), renowned for its dense concentration of private educational institutions. The second focal point (No. 2 in Figure B\ref{fig:clusterspace}) is another downtown area (Type B) that attracts a substantial number of non-Seoul residents, including tourists. This unexpected finding can be attributed to the nature of type A areas, which are specialized zones attracting visitors from beyond Seoul's borders for specific industry-related shopping, such as digital, clothing, and large retail complexes as mentioned before. Our study's focus on Seoul residents' consumption activities meant that shopping trips from outside the city's limits were beyond our analytical scope. 

In contrast to the No. 2 cluster in Type B, which stands as a prime commercial destination attracting a wide array of people, including tourists, the No. 1 cluster within the same type serves as another focal point, particularly for Seoul residents' long-distance trips. Given South Korea's strong emphasis on education—especially in Seoul, a city leading in private educational services—these areas draw individuals from across the city and beyond. Parents accompanying their children to educational institutions often engage in shopping in the nearby areas, thereby frequently contributing to these locations' status as long-distance shopping hubs. The observation that many shopping trips headed towards this cluster originate from Types C or D, known for their high residential population densities, reinforces this inference.

Interestingly, the pandemic has seemed to affect the shopping dynamics in the two downtown commercial areas previously identified as long-distance shopping focal points in a different way. Both areas experienced a contraction in activity levels, with commercial area 2 suffering a more pronounced decline. In addition, by 2023, commercial area 1 began showing signs of recovery, with a more concentrated pattern of shopping activity, unlike commercial area 2, which did not exhibit significant recovery signs. It seems to have contracted further, instead. 



\setcounter{table}{0}
\setcounter{figure}{0} 
\renewcommand{\thetable}{A\arabic{table}}
\renewcommand{\thefigure}{A\arabic{figure}}
\end{appendices}
\end{document}